\documentclass[prl,letterpaper,amsfonts]{revtex4}
\usepackage{color}
\usepackage{graphicx}

\usepackage{amsthm}
\usepackage{amsmath}
\usepackage{amsfonts}

\usepackage{amssymb}

\usepackage{bm}
 \usepackage{lineno}

\begin{document}

\bibliographystyle{apsrev}

% Dirac Bra
\newcommand{\bra}[1]{\ensuremath{\left \langle #1 \right |}}
% Dirac Ket
\newcommand{\ket}[1]{\ensuremath{\left | #1 \right \rangle}}
% Inner Product
\newcommand{\braket}[2]{\ensuremath{\left \langle #1 \right | \left. #2 \right \rangle}}
% Trace
\newcommand{\tr}{\ensuremath{\mbox{Tr}}}
\newcommand{\R}{{\bf r}H}
\newcommand{\V}{\widehat{{\bf r}H}}
\newcommand{\U}{{\bf t}H}
% Hilbert space
\renewcommand{\H}{\ensuremath{\mathcal{H}}}
% Variance
\newcommand{\var}{\ensuremath{ \mbox{var}}}
% Sample variance
\newcommand{\hvar}[1]{\ensuremath{ \hat{\mbox{var}} \left ( #1 \right )}}
\newcommand\Prefix[3]{\vphantom{#3}#1#2#3}

\title{How modular structure can simplify tasks on networks}
\author{Binh-Minh Bui-Xuan and Nick S. Jones}

%\affiliation{Laboratoire d'Informatique de Paris 6, Universit\'e Pierre et Marie Curie}
\affiliation{CNRS -- APR, LIP6, UPMC}

\affiliation{Department of Mathematics, Imperial College London}

%\date{\today}

\begin{abstract}
By considering the  task of finding the shortest walk through a network we find an algorithm for which the run time is not as $O(2^n)$, with $n$ being the number of nodes, but instead scales with the number of nodes in a coarsened network. This  coarsened network has a number of nodes related to the number of dense regions in the original graph. %Hamiltonian walk can be viewed as the task of minimizing the energy of a walk on a graph which has an energy cost associated with length.
Since we exploit a form of local community detection as a preprocessing, this work gives support to the project of developing heuristic algorithms for detecting dense regions in networks: preprocessing of this kind can accelerate optimization tasks on networks. Our work also suggests a class of empirical conjectures for how structural features of efficient networked systems might scale with system size.
\end{abstract}

\maketitle
  %\linenumbers
The last decade has seen a widespread appreciation that networks in Nature have a structure which makes them poorly modeled as samples from the more traditional random graph ensembles \cite{Erdos59,Newman10}. A feature of many real networks is that a marked modular (or community) structure is present \cite{Porter09,Lancichinetti10,Fenn10,Agarwal12} where a network community is a subset of nodes with relatively dense connections within the subset but sparse connections to the rest of the network. Though algorithms for detecting communities are actively constructed in physics, mathematics, engineering and computer science, %\cite{Reichardt04}, %slightly changed
the reason for detecting these dense regions is not always articulated. Studies of empirical graphs suggest that nodes in the same community tend to have similar properties in the world and thus community detection can help us, for example, assign functional labels to uncharacterized nodes \cite{Porter09}. In this paper we investigate how community detection can also help simplify problems on graphs.

Even though community detection tasks can be hard \cite{Brandes,Newman04} experience with greedy algorithms suggests that plausible solutions can be found quickly for networks with a pronounced community structure (though, of course, sub-optimal solutions need to be treated with care \cite{Good10}) \cite{Porter09}. Recent theoretical work in network physics and computer science also suggests that for certain types of graphs, community detection need not be costly \cite{perscomm10,Moore11,Newman12}. It is also the case that a marked community structure is present in many empirical networks and that networks with similar functions appear to have similar community structure \cite{Lancichinetti10,Fenn10,Agarwal12}. It is possible that this similarity occurs because community structure constrains dynamics on the graphs. Indeed it has been found that particular choices of dynamics on networks can in turn correspond to particular methods for detecting communities in graphs \cite{Lambiotte08}. %Given that the modular structure of networks appears to  constrain strongly the dynamics that occur on them, it seems reasonable to ask whether this common feature will also simplify optimization tasks on the graphs. %...In particular, just as there is an association between dynamics on a graph and natural choices for communities, it might be the case that certain optimization problems are simple on certain graphs...
The literature asking why the networks we observe in the world are modular is substantial \cite{Simon62}: one might speculate that assembly rules for real networks are such as to simplify either optimization tasks or dynamics on them and so in turn this leads to a pronounced community structure. %!! small change
The notion that networks in nature  are often optimized for transport is an established part of theoretical biological physics (e.g. \cite{West97}).

\vspace{3mm}

\emph{Parameterized Complexity:} Weakly coupled to the stream of empirically motivated networks literature is recent work in computational complexity called parameterized complexity, or fixed parameter tractability \cite{Niedermeier06}. The concerns of this vibrant field are common to the empirical interests of the network research community: how do (parameterized) constraints on graph structure simplify problems on graphs? Researchers in empirical studies of networks ask: how are real world networks simple? Asking whether some hard problems are simple on empirical graphs is thus natural. In the following, though we use recent work from parameterized complexity, we will not be providing algorithms that have computational cost scaling in polynomial time with network size; instead we will show that a problem scaling like $2^n$ ($n$ is the number of nodes in the graph) can be converted into one which scales like $\leq ~2^{\tilde{n}}$  but where we can suggest $\tilde{n}\ll n$ by relating $\tilde{n}$ to (a local version of) the number of communities in the graph.

\vspace{3mm}

%Our motivation:

\emph{Hamiltonian Walk and Communities} Motivated by the above observations, we ask whether a particular problem, Hamiltonian Walk, HAMWALK (an NP-hard task \cite{note1}) is simpler on networks with pronounced community structure. We define a Hamiltonian Walk as a shortest closed walk on a graph which visits every node at least once \cite{Goodman74,note0}. The study of self avoiding walks on lattices and fractals is an established area of probability and statistical physics and modifications which allow limited self crossings have been considered \cite{Madras96}; problems of this kind are of broad relevance to understanding percolation and polymer phenomena. We note, of course, that the interface between problems in computational complexity and statistical physics is now a lively one \cite{Mezard09}.

\begin{figure}[h!]
\includegraphics[angle = 0, width = 12cm, %, height=5cm %%@
keepaspectratio=true
]{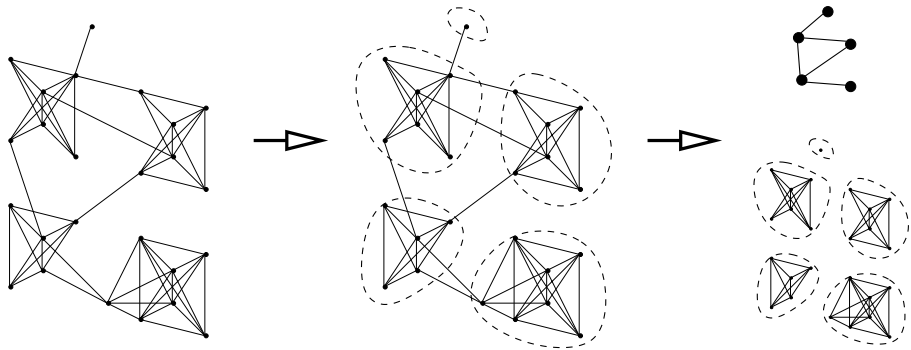}%fourhearty4.eps}%threeheartstest.eps}
\caption{\label{Fig 1.}
Community detection can be fast if pronounced communities are present \cite{perscomm10,Moore11}. Solutions to optimization problems on the coarsened graph can sometimes be converted into solutions for the full graph.}
\end{figure}

We hypothesize that partitioning graphs into communities, coarsening the graphs to yield smaller graphs with nodes representing entire communities \cite{Radicchi08}, and then solving problems on the coarsened graph and on the individual communities of the full graph in combination, might lead to significant computational speed-ups for some real graphs and appropriately chosen optimization tasks (see Fig \ref{Fig 1.}). {\color{black} We hope to exploit the fact, noted above, that empirical networks often have pronounced community structure and finding this structure need not be hard.}

For real networks this hypothesis could be converted into a class of heuristic algorithms for solving optimization problems.  It thus seems desirable to see whether the preceding intuition can be expressed mathematically: is there a class of (crudely) realistic graphs for which HAMWALK has a runtime scaling which is provably less than the $O(2^n)$ bound that can currently be achieved \cite{note1}? We believe this both motivates further work in providing  tighter runtime bounds for more realistic graphs, justifies the development of appropriate heuristic optimization algorithms and makes a connection between parameterized complexity and network empirics: making explicit the notion that the modular network structure we observe in Nature could help simplify tasks for networked systems.

\vspace{3mm}

\emph{A local clustering algorithm:} In Ref. \cite{perscomm10} the authors study the runtime of finding partitions of networks into clusters (disjoint sets of nodes) where each cluster, $C_i$  (for all $i$), has (i) a total number of links connected to nodes not in $C_i$ that is $\leq\delta$ (call this the degree of $C_i$) and (ii) the number of pairs of nodes in $C_i$ between which there is no link is $\leq\mu$. This bears some resemblance to the local community detection in Ref. \cite{clauset05} and seeks to identify sets of densely connected nodes ($\mu$ small) which are isolated from other such sets ($\delta$ small). If $\delta$ is treated as a fixed input then, remarkably, this problem can be solved in randomized time $2^{O(\mu)}n^{O(1)}$. Similarly if $\mu$ is held fixed then the run time is $2^{O(\delta)}n^{O(1)}$ \cite{perscomm10}.

%Counting the number of paths that visit all nodes exactly once (called Hamiltonian Path) is

%\emph{Definition of HAMWALK and a simple motivating class of graphs}
%Note that HAMPATH is what many physicists call HAMWALK.
%
%"Given any connected graph G, it is possible to start at an
%arbitrary point u of G, walk in some sequence along the lines of G and return to
%the starting point u having passed through every point in G at least once. In
%general, such a walk might pass through some points, and traverse some lines,
%more than once. We call such a walk a closed spanning walk of G. A Hamiltonian
%walk in G is a closed spanning walk of minimum length." This is a quote from Goodman

\vspace{3mm}

\emph{Special cases:} We first run an algorithm that detects all clusters in graph, $G$, with degree $\leq \delta$ and $\leq \mu$ missing links \cite{perscomm10}. In the simple case when $\delta=2$ and $\mu=0$ a naive solution is as follows. Define a coarsened version of $G$, $G'$, in which, for each cluster taken consecutively (the order is irrelevant), all nodes of the cluster are removed and substituted for a single node, called a cluster-node, which is connected to the nodes ($\leq 2$ nodes) to which the cluster was previously connected. We then solve HAMWALK on $G'$ using the Held-Karp algorithm a \cite{note1,Held62} and obtain a walk (in $G'$) as result. We finally expand this walk to become a walk in $G$ simply by a replacement of every cluster-node by its original cluster of $G$ and quickly computing an appropriate paths through the clique (note that $\mu=0$ so this task is simple). It is a standard exercise to check the obtained walk is indeed a Hamiltonian walk of $G$. The intuition behind is that sets of nodes which are strongly isolated from the rest of the graph (like in the case of $\delta=2$) allow marked simplifications of problems on the graph. This provokes the question that we consider in the following: does this intuition hold for richer classes of graphs, with more general $\delta$ and $\mu$?

\begin{figure}[h!]
\includegraphics[angle = 0, width = 17cm, %, height=5cm %%@
keepaspectratio=true
]{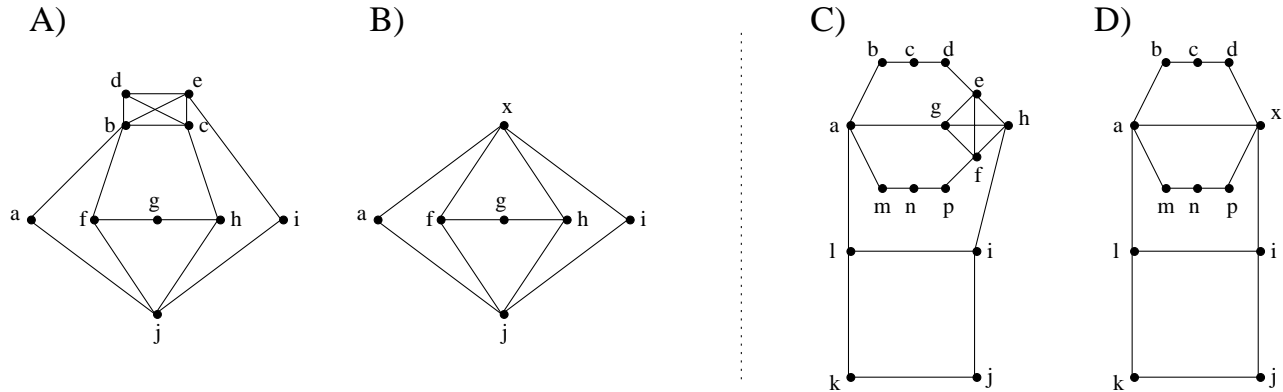}
\caption{\label{Fig 2.}
  \emph{A}) Graph $G$, where $\{b,c,d,e\}$ is a clique.
  \emph{B}) Coarsened graph $G'$ where clique $\{b,c,d,e\}$ is replaced by cluster-node $x$.
\emph{C}) and \emph{D}) a further example of original graph $G$ and coarsened graph $G'$.
}
\end{figure}

Unfortunately, unlike the previous case, not all solutions to HAMWALK($G'$) can be easily modified to make solutions to HAMWALK($G$).
Consider the graph $G$ and its coarsened graph $G'$ in Fig.~\ref{Fig 2.} A) and B).
Both $W'_1=(a,x,f,g,h,x,i,j,a)$ and $W'_2=(a,j,f,g,h,j,i,x,a)$ are Hamiltonian walks of $G'$.
Here, $W'_1$ can be expandable to a Hamiltonian walk of $G$ just by replacing the first occurrence of $x$ with $b$ and the second occurrence of $x$ with $c,d,e$, namely to obtain $W_1=(a,b,f,g,h,c,d,e,i,j,a)$ (a cycle, hence optimal in size).
However, applying such local expansions on $W'_2$ will not be as successful (not a cycle, because of multiple occurrence of $j$, %nj - this previously said $a$. I presume this is a typo?
hence, longer walk than before).

Given the above one might conclude that if we carefully count the number of times that each cluster-node is visited, that will help us extend, simply, our walks on $G'$ to walks on $G$.
In the preceding example $W'_1$ visits $x$ twice, whereas $W'_2$ visits $x$ only once, so it could be that the number of visits to the cluster-node $x$ is important.
However, graphs exist which allow two solutions to HAMWALK($G'$) which visit cluster-nodes the same number of times but which do not both allow a simple expansion to a solution for $G$ {\sl e.g.}, Fig.~\ref{Fig 2.} C) and D). In $G'$ (Fig.~\ref{Fig 2.}D), both $W'_1=(a,b,c,d,x,i,j,k,l,a,m,n,p,x,a)$ and $W'_2=(a,b,c,d,x,i,j,k,l,i,x,p,n,m,a)$ are walks of $G'$ visiting cluster-node $x$ twice.
$W'_1$ can be locally expanded to a Hamiltonian walk of $G$ by replacing the first occurrence of $x$ with $e,h$ and the second occurrence of $x$ with $f,g$:  $W_1=(a,b,c,d,e,h,i,j,k,l,a,m,n,p,f,g,a)$ is a Hamiltonian walk of $G$.
One can check that $W'_2$ cannot be expanded to a Hamiltonian walk of $G$ only by replacing occurrences of $x$ with vertices from $\{e,f,g,h\}$, by exhaustive case analysis for instance (the edge $(h,i)$ must be traversed twice).

%\begin{figure}[h!]
%\includegraphics[angle = 0, width = 5cm, %, height=5cm %%@
%keepaspectratio=true
%]{ex2.eps}
%\caption{\label{Fig 3.}
%}
%\end{figure}

\emph{Solving HAMWALK with parameters $\delta$ and $\mu$} We now consider the case when $\delta$ and $\mu$ are unspecified positive integers, we proceed as follows.
Each $i^{\rm{th}}$ cluster, $C_i$, will contain a set of shell nodes, $S_i$, which are defined as being in $C_i$ and either have a link to a node which is not in $C_i$, or lack a link to a node which is in $C_i$.
When $|C_i|>2\cdot|S_i|$, we define another set of nodes, called second shell nodes, $T_i$, which are taken randomly from $C_i\setminus S_i$ so that  $|S_i|=|T_i|$ ($|S_i|\leq \delta + 2\mu$).
We call nodes which are in each detected cluster but which are neither shell nodes nor second shell nodes, good bulk nodes, $GB_i=C_i\setminus S_i \setminus T_i$. $GB_i$ is thus a clique; it is this simple structure we will exploit in the following. The set $T_i$ is a device which will help simplify our proof.
%This way, we have in particular that $GB_i$ is a clique, the only way to visit $GB_i$ from outside $C_i$ is passing through $S_i$, and, finally, any node in $S_i\cup T_i$ can always connect to a node in $T_i$.

We now define a coarsened version of $G$, $G'$, in which, for each cluster with degree $\leq \delta$, and at most $\mu$ missing links and $|C_i|>2\cdot|S_i|$, all nodes in the good bulk are removed and substituted for a single node, called the coarsened good bulk node, $b_i$, which is connected to all of the shell and second shell nodes in the cluster. We will later discuss coarsened walks: a walk on $G$ is coarsened to a walk on $G'$ by identifying consecutive (or single) walker visits to nodes in $GB_i$ and substituting them for single visits to the coarsened good bulk node $b_i$.

{\color{black} The outline of our proof is as follows: we create a coarsened graph, $G'$, in which all nodes in each clique $GB_i$ are represented as single node $b_i$ and all other nodes are left untouched. We solve HAMWALK by a standard method on the coarsened graph. We show that, because of the simplicity of the cliques $GB_i$ and, noting the properties of the shell and second shell, that this walk can be converted into a solution to HAMWALK on the original graph in a time polynomial in $n$.} The use of the two shells will allow us to avoid problems identified in the examples above.

\vspace{3mm}

\emph{Claim 1:} any Hamiltonian walk of $G'$ can, with resources polynomial in network size, be converted to a Hamiltonian walk of $G'$ that visits every coarsened good bulk node once and only once.

\vspace{3mm}

\emph{Proof of Claim 1:} It is easy to check whether a Hamiltonian walk of $G'$ visits every coarsened good bulk node once and only once.
If this is not the case, we can repeatedly perform the following substitutions.
Call $b$ one of the coarsened good bulk nodes which is visited more than once.
Consider any {\color{black} walk with a} visit to $b$: it is either of the form
a) $s_ibs_i$ or $s_ibs_j$ $i\neq j$, where $s_i$ and $s_j$ are nodes in the shell of the cluster to which $b$ belongs,
or b) $t_ibt_i$ or $t_ibt_j$ $i\neq j$, where $t_i$ and $t_j$ are second shell nodes,
or c) $s_ibt_j$ (or $t_ibs_j$), where $s_i$ is a shell node and $t_j$ a second shell node.
Now we can obtain another closed walk that still visits every node of $G'$ at least once, $b$ included, as follows:
a) replace $s_ibs_i$ by $s_i$, or replace $s_ibs_j$ by $s_its_j$, where $t$ is any node from the second shell;
b) replace $t_ibt_i$ by  $t_i$, or replace $t_ibt_j$ by $t_it_j$;
c) replace $s_ibt_j$ by $s_it_j$.
In all cases, the length of the modified walk is not increased, that is, if the original is a Hamiltonian walk then the modified walk is still a Hamiltonian walk of $G'$. $\Box$

Given any Hamiltonian walk of $G'$ we can thus obtain a Hamiltonian walk of $G'$ having the property described in \emph{Claim 1}.
Denote its length by $w_{G'}$. We can un-coarsen the walk into a closed walk on $G$ by locally extending using a greedy approach at every good bulk node.
What \emph{Claim 1} implies is that the extended walk has length $w_{G'}+\sum_i(|GB_i|-1)$ (by exploiting the clique structure of $GB_i$).
Clearly, $w_{G'}+\sum_i(|GB_i|-1) \geq w_G$, where $w_G$ is the length of Hamiltonian walks of $G$.
%Combining the inequalities, we deduce that locally extending the Hamiltonian walk of $G'$ given by \emph{ Claim 1} will result in a Hamiltonian walk of $G$, that is,
Now, if this were an equality, we would have shown the following.

\vspace{3mm}

\emph{Main claim:} any solution to HAMWALK($G'$) can, with resources polynomial in network size, be converted into solutions of HAMWALK($G$).

\vspace{3mm}

\emph{Claim 2:} there is always a Hamiltonian walk of $G$ which, when coarsened to a walk on $G'$, visits coarsened good bulk nodes once and only once.

\vspace{3mm}

The \emph{Main claim} is true provided that \emph{Claim 2} is true.
This is because it would imply that $w_{G'}+\sum_i(|GB_i|-1)$ is equal to $w_G$
(pick the Hamiltonian walk given by \emph{Claim 2};
consider its coarsening on $G'$ whose length has to be $w_G-\sum_i(|GB_i|-1)$;
but this is also a closed walk of $G'$ that visits every node of $G'$ at least once and thus is of length $\geq w_{G'}$.  We know from \emph{Claim 1} that $w_{G'}+\sum_i(|GB_i|-1) \geq w_G$ and hence the \emph{Main claim} follows).

\vspace{3mm}

\emph{Proof of Claim 2:} Clusters are pairwise independent, so that we will only give a proof for \emph{Claim 2} with respect to a particular cluster $C$, with set of shell nodes $S$, set of second shell nodes $T$, and set of good bulk nodes $GB$.
Among the Hamiltonian walks of $G$, consider one that makes distinct visits to nodes in $GB$ the least number of times (a distinct visit to $GB$ is a part of a walk containing a contiguous sequence of nodes in $GB$ immediately preceded and followed by a visit to nodes not in $GB$).
By contradiction suppose this walk contains repeated distinct visits to $GB$.
%The Hamiltonian walk we consider will either visit $T\cup GB$
%a) only once
%or b) at least $p\geq2$ times.
Consider visits to $T\cup GB$: because nodes outside $C$ can connect to $T\cup GB$ only via the shell $S$, such a visit has to be of the form $s_iws_j$, where $s_i$ and $s_j$ are shell nodes and $w$ a walk in $T\cup GB$.
Denote by $p$ the number of times $T\cup GB$ is visited: some thought shows that $p$ is no greater than $|S|$ and hence not greater than $|T|$.
Now replace the first $p-1$ visits to $T\cup GB$ keeping the same shell nodes but the walk in $T\cup GB$ is substituted for a random (chosen without reuse) second shell node: for instance, $s_iws_j$ is substituted for $s_it_ks_j$ with some $t_k\in T$.
Denote the last visit to $T\cup GB$ by $s_kus_l$ {\color{black} (this includes the $p=1$ case)} and substitute it for $s_kvs_l$ where $v$ contains all nodes in $T\cup GB$ (except for those of $T$ that have been used for the $p-1$ previous visits) {\color{black} and $v$ visits all nodes of $GB$ contiguously}.
%Moreover, we arrange $v$ so that nodes in $GB$ are visited consecutively.
We now consider the result of these substitutions: clearly, this is still a closed walk that visits every node of $G$ at least once, furthermore, its has the same length {\color{black} (or less)} as the original walk before substitutions.
In other words, we have obtained a Hamiltonian walk of $G$, but one that visits nodes in $GB$ only once: a
contradiction. $\Box$

\vspace{3mm}

\emph{Algorithm solving HAMWALK:} given $\mu$ and $\delta$ (where either $\mu$ or $\delta$, but not both, might be a function of $n$) take $G$ on $n$ nodes and create the graph $G'$ using \cite{perscomm10}. In the case that $\mu$ is fixed this takes a time $n^{O(1)}2^{O(\delta)}$ \cite{perscomm10}. Calculate the matrix of shortest paths for $G'$, and use the Held-Karp algorithm to find a solution to HAMWALK($G'$) \cite{note1,Held62}. Convert it into one having the property described in \emph{Claim 1}. Locally extend this to a walk in $G$ using a greedy substitution at every {\color{black} coarsened} good bulk node. Return the latter as a solution to HAMWALK($G$).
Runtime:
$~2^{(2\delta+4\mu +1) n_c+n'}+ n^{O(1)}2^{O(\delta)}$ where $n_c$ is the number of clusters with degree $\leq \delta$, at most $\mu$ missing links, and $|GB|>2\cdot|S|$ and $n'$ is the number of nodes which are not in clusters with $|GB|>2\cdot|S|$ (note $|S|\leq \delta+2\mu$). $\Box$

\vspace{3mm}

\emph{A note on Clique-width:} Clique-width is a quantity which has proved very popular in the area of parameterized complexity: it is related to a natural approach to assembling a graph \cite{courcelle00}. We first define clique-width and then note some implications of our work (thereby illustrating its relevance in parameterized complexity). We suppose a $k$-graph has nodes with labels from the set $\{1, 2, . . . , k\}$. We define a seed $k$-graph with one node and a label from this set. The clique-width of a graph $G$ is the smallest integer $k$ such that $G$ can be composed by repeated use of four simple operations: a) generate: make a seed $k$-graph labeled by $i$, b) disjoint union: two distinct graphs are treated as disconnected components of the same graph, c) combine: linking all nodes with label $j$ with nodes labeled $i$, d) relabel: all indices $i$ replaced with $j$. This protocol can generate graphs which contain cliques. In this paper, clique-width can be connected to the case where there are no missing links inside the cliques identified ($\mu=0$). In this case, it can be proved fairly easily that the clique-width of the graph is $\leq k=(\delta n_c+n')$ and a corresponding protocol for constructing the graph using the above four operations can be provided (this witness protocol, demonstrating that the bound can be met, is called a $k$-expression). Given this protocol a number of Monadic Second Order Logic (MSOL) problems can be solved (see \cite{Courcelle} and the numerous MSOL problems therein) including the classic problem `Minimum Dominating Set' \cite{Garey} in a time $2^{O(\delta n_c+n')}$. This last follows from the recent observation that, given a $k$-expression for clique-width $k$, Minimum Dominating Set can be solved in a time $2^{2 k}$ \cite{Bodlaender}. Thus our result for HAMWALK can be extended to other problems.

%A few things -- I guess it may be a bit computer sciencey for PRL as it stands? A bit less caution about the scientific utility and some bio/eng/etc examples of important optimisation tasks on structured network may frame it better?

\vspace{3mm}

\emph{Discussion and Conclusions:}
{\color{black} We have proved that it is possible to solve HAMWALK
 in time $~2^{(2\delta+4\mu +1) n_c+n'} + n^{O(1)}2^{O(\delta)}$. %(in the setting $\mu$ constant and $f(\delta) = O(\delta)$).
 Some care is required in the interpretation of this runtime. While $\delta$ and $\mu$ are parameters of the algorithm (with forms which can be specified independent of the graph) by contrast $n_c$ is a feature of the graph for a given $\delta$ and $\mu$. Despite this, one can hope to construct graphs which have a given, $n_c$, $n'$, $\delta$ and $\mu$.} To help the interpretation of this result we thus consider the following graph family: all nodes are inside one and only one of $n_c$ clusters such that for each cluster, $\mu$ is constant, {\color{black}$2^{O(\delta)}=  n^{O(1)}$}, $n_c\delta\leq O(1) \log n$ (for this family $n'=0$) and each cluster is of size $|C|> 2(\delta+2\mu)$. This is a graph family composed of dense clusters in which links between the clusters and missing links inside the clusters are proportionately rare (an equivalent family with the roles of $\mu$ and $\delta$ reversed could also be considered): as graphs increase in size the number of clusters is relatively slow growing. HAMWALK can be solved on this family in time $n^{O(1)}$ (if the values of $\mu$ and $\delta$ are not known in advance, finding appropriate choices only yields a polynomial time overhead). In the intuitive case with $\mu$ and $\delta$ both constant, the number of clusters increases as the logarithm of the number of nodes.
 {\color{black}Our abstract complexity based argument can be used to inspire a class of empirical conjectures about optimal growing networks in Nature.} Suppose that, in order to thrive, a network in the world has to solve HAMWALK on itself efficiently for larger and larger system sizes.
  Given the above, we might thus hypothesize that, if we have observations of the system at a variety of network sizes, we will find that the number of communities would increase like the logarithm of the number of nodes. While few natural systems are likely optimized to solve HAMWALK, this form of reasoning might allow us to relate the system-size scaling of structural features to the tasks networked-systems are optimized to solve. Why natural networks might show modular structure is a canonical question; beyond conjectured roles for communities like helping networked systems to control their dynamics, or to be more evolvable \cite{Simon62}, we have provided a concrete setting in which community structure helps simplify optimization tasks.% small change
%\vspace{-2mm}

The family specified above was selected to share some similarities with real networks, while remaining mathematically tractable: many real networks do have sets of densely connected nodes which are relatively isolated from the rest of the network. The bound, $\delta$, gives some indication of the modular nature of the graph: for graphs with appropriately small $\delta$ {\color{black} the clusters are more isolated from each other and it is easier to solve HAMWALK}; for appropriately small $\mu$ again HAMWALK is easier to solve.  If $n'\ll n$, indicating a graph with many nodes inside clusters, then again it is much easier to solve HAMWALK.  The bound we proved is by construction: it might be possible to solve HAMWALK in faster time. We obtained our bound on HAMWALK by coarsening the graph on $n$ nodes and solving problems on the reduced version. It is known that empirical networks have hierarchical community structure: dense regions embedded inside others. Authors have considered renormalization approaches to such networks \cite{Radicchi08}. How this hierarchical structure simplifies problems on graphs, and how repeated coarsenings of the full graph might help is open. %We believe that these ideas can be used to motivate heuristic algorithms for real graphs which can then be compared to the canon of existing approaches.
In this setting, not only does modular structure constrain dynamics on graphs but it can also simplify problems on graphs. %**Given a choice of
Many practical optimization problems are posed on real graphs with modular structure. The preceding gives a hint that some of the optimization problems we are interested in might allow heuristics which run in time scaling with the number of communities (and parameterizations of their isolation) rather than the number of nodes: this gives a strong justification for the exercise of developing community detection algorithms. %**Since networks might have evolved to be efficient wrt to optimization tasks this can be interpreted as an empirical claim about the world.**

\vspace{3mm}

{\emph{Thanks to:} Sumeet Agarwal and also to Sam Johnson, Iain Johnston and Simone Severini }


\begin{thebibliography}{30}
\bibitem{note1} It is straightforward to recast HAMWALK as a travelling salesman problem which can be solved in time $O(2^n)$ \cite{Held62}.
\bibitem{note0} A Hamiltonian Walk sometimes also means a covering walk which is strictly self avoiding.
\end{thebibliography}
\end{document}